\documentclass[%
english,aps,prr,reprint,
nobibnotes,footinbib,
amsmath,amssymb,
superscriptaddress
]{revtex4-2}

\usepackage[utf8]{inputenc}
\usepackage[T1]{fontenc}
\usepackage{amsmath, amssymb, physics, braket}
\usepackage{siunitx}

\usepackage{tikz}
\usepackage{graphicx}
\usepackage{hyperref}

\newcommand{\x}{\Delta/\Omega}
\newcommand{\y}{\mathcal{R}_\text{b}}

\newcommand{\Ha}{\mathcal{H}}
\newcommand{\n}{\hat{n}}
\newcommand{\s}{\hat{\sigma}^x}
\newcommand{\B}{\bar{\beta}}
\newcommand{\D}{\xi\Delta q}

\setcitestyle{super}

\usetikzlibrary{shapes.geometric, positioning}
\usepackage{tikz}

\definecolor{Beige}{RGB}{251, 241, 199}
\definecolor{Orange}{RGB}{255, 158, 74}
\definecolor{Blue}{RGB}{114, 158, 206}
\definecolor{Red}{RGB}{237, 102, 93}

\tikzstyle{o}=[circle, draw, fill=Beige, minimum size=15, inner sep=0]

\tikzstyle{n}=[circle, draw, fill=gray, minimum size=7, inner sep=0]

\tikzstyle{ln}=[isosceles triangle, isosceles triangle apex angle=70, draw, fill=gray, minimum size=6, inner sep=0]
\tikzstyle{rn}=[isosceles triangle, isosceles triangle apex angle=70, rotate=180, draw, fill=gray, minimum size=6, inner sep=0]

\tikzstyle{p}=[black, thick]

\newcommand{\HuseFisher}{
	\begin{tabular}{cc}
		$\delta = 0$ & $0 < \delta < \delta_L$ \\
		Conformal $(\nu > \B)$ & Chiral $(\nu = \B)$ \\
		\begin{tikzpicture}[baseline=(current bounding box.center)]

			\shade[top color=white, bottom color=Blue] (0, 0) rectangle (1.2, 0.15);

			\draw[p, draw=Blue] (1.2, 0) parabola (2.8, 1.8);
			\draw[p, draw=Red] (0, 1.5) .. controls (0.6, 1) and (1, 0.6) .. (1.2, 0);
			\draw[p, draw=Red] (2.4, 1.8) .. controls (1.8, 1.3) and (1.5, 0.9) .. (1.2, 0);

			\draw[p, ->] (0, 0) -- (0, 2);
			\draw[p, ->] (0, 0) -- (3, 0);
			\draw (2.2, 0) -- node[below]{$T$} (3, 0);
			\draw (0, 1.5) -- node[left]{${\color{Red}\frac{1}{\xi}}, {\color{Blue}q}$} (0, 2);

			\draw[thick, dashed] (1.2, 0) --  ++(0, 2);
			\draw (1.2, 1.8) -- node[left]{C} ++(0, 0);
			\draw (1.2, 1.8) -- node[right]{IC} ++(0, 0);

			\draw (0, 0) -- node[left]{${\color{Red} 0}, {\color{Blue} \frac{1}{p}}$} (0, 0);

		\end{tikzpicture} &
		\begin{tikzpicture}[baseline=(current bounding box.center)]

			\shade[top color=white, bottom color=Blue] (0, 0) rectangle (1.2, 0.15);

			\draw[p, draw=Red] (0, 1.5) .. controls (0.6, 1) and (1, 0.6) .. (1.2, 0);
			\draw[p, draw=Red] (2.4, 1.8) .. controls (1.8, 1.3) and (1.5, 0.9) .. (1.2, 0);
			\draw[p, draw=Blue] (2.6, 1.6) .. controls (1.9, 1.2) and (1.6, 1.05) .. (1.2, 0);

			\draw[p, ->] (0, 0) -- (0, 2);
			\draw[p, ->] (0, 0) -- (3, 0);
			\draw (2.2, 0) -- node[below]{$T$} (3, 0);
			\draw (0, 1.5) -- node[left]{${\color{Red}\frac{1}{\xi}}, {\color{Blue}q}$} (0, 2);

			\draw[thick, dashed] (1.2, 0) --  ++(0, 2);
			\draw (1.2, 1.8) -- node[left]{C} ++(0, 0);
			\draw (1.2, 1.8) -- node[right]{IC} ++(0, 0);

			\draw (0, 0) -- node[left]{${\color{Red} 0}, {\color{Blue} \frac{1}{p}}$} (0, 0);

		\end{tikzpicture}
	\end{tabular}
	\]
	\[
	\begin{tabular}{c}
		$\delta > \delta_L$ \\
		PT - Floating - KT \\
		\begin{tikzpicture}[baseline=(current bounding box.center)]

			\shade[top color=white, bottom color=Blue] (0, 0) rectangle (1, 0.15);
			\shade[top color=white, bottom color=Red] (1, 0) rectangle (1.7, 0.15);

			\draw[p, draw=Blue] (1, 0) arc (180:95:1.7);
			\draw[p, draw=Red] (1, 0) arc (0:80:1.2);
			\draw[p, draw=Red] (2.9, 1.5) .. controls (1.95, 1) and (2, 0) .. (1.7, 0);

			\draw[p, ->] (0, 0) -- (0, 2);
			\draw[p, ->] (0, 0) -- (3, 0);
			\draw (2.2, 0) -- node[below]{$T$} (3, 0);
			\draw (0, 1.5) -- node[left]{${\color{Red}\frac{1}{\xi}}, {\color{Blue}q}$} (0, 2);

			\draw[thick, dashed] (1, 0) --  ++(0, 2);
			\draw (1, 1.8) -- node[left]{C} ++(0, 0);
			\draw (1, 1.8) -- node[right]{FL} ++(0, 0);
			\draw[thick, dashed] (1.7, 0) --  ++(0, 2);
			\draw (1.7, 1.8) -- node[right]{IC} ++(0, 0);

			\draw (0, 0) -- node[left]{${\color{Red} 0}, {\color{Blue} \frac{1}{p}}$} (0, 0);

		\end{tikzpicture}
	\end{tabular}
}

\begin{document}

\author{Ivo A. Maceira}
\affiliation{Institute of Physics, École Polytechnique Fédérale de Lausanne (EPFL), CH-1015 Lausanne, Switzerland.}

\author{Natalia Chepiga}
\affiliation{Department of Quantum Nanoscience, Kavli Institute of Nanoscience, Delft University of Technology, Lorentzweg 1, 2628 CJ Delft, The Netherlands.}

\author{Frédéric Mila}
\affiliation{Institute of Physics, École Polytechnique Fédérale de Lausanne (EPFL), CH-1015 Lausanne, Switzerland.}

\date{\today}


\title{Conformal and chiral phase transitions in Rydberg chains}

\begin{abstract}
Using density matrix renormalization group simulations on open chains, we map out the wave-vector in the incommensurate disordered phase of a realistic model of Rydberg chains with $1/r^6$ interactions, and we locate and characterize the points along the commensurate lines where the transition out of the period 3 and 4 phases is conformal. We confirm that it is 3-state Potts for the period-3 phase, and we show that it is Ashkin-Teller with $\nu\simeq 0.80$ for the period-4 phase. We further show that close to these points the transition is still continuous, but with a completely different scaling of the wave-vector, in agreement with a chiral transition. Finally, we propose to use the conformal points as benchmarks for Kibble-Zurek experiments, defining a roadmap towards a conclusive identification of the chiral universality class.
\end{abstract}

\maketitle

\section{Introduction}
\label{sec:introduction}

In recent experiments on chains of Rydberg atoms with programmable interactions\cite{Bernien.Schwartz.ea:2017, Keesling.Omran.ea:2019}, quantum phase transitions between commensurate (C) ordered phases of periods $p=3,4$ and an incommensurate (IC) disordered phase were probed dynamically using the quantum Kibble-Zurek mechanism\cite{Kibble:1976, Zurek:1985, Zurek.Dorner.ea:2005}. These experiments have renewed the interest in the problem of IC-C transitions first studied in the 80's and 90's in the context of adsorbed monolayers\cite{Abernathy.Song.ea:1994, Schreiner.Jacobi.ea:1994}. The IC-C critical behavior of a minimal model introduced to describe such transitions, the $p$-state chiral clock model\cite{Ostlund:1981, Huse:1981}, contains most of the relevant physics.

The IC-C transition with $p\geq5$ happens through an intermediate gapless phase of central charge $c=1$ characterized by incommensurate correlations. The dominant wave-vector $q$ is not frozen to any specific value but changes continuously - floats - through the phase, therefore referred to as a \textit{floating} phase\cite{Elitzur.Pearson.ea:1979, Cardy:1980, Ostlund:1981}. The disorder to floating transition is in the Kosterlitz-Thouless (KT) universality class\cite{Kosterlitz.Thouless:1973}, with exponentially diverging correlation length $\xi$. One reaches the ordered phase through a Pokrovsky-Talapov \cite{Pokrovsky.Talapov:1979} (PT) transition where the wave-vector $q$ (which we define in units of $2\pi$) goes to $1/p$ as a power-law with the exponent $\B=1/2=\nu$, where $\nu$ is the correlation length critical exponent.

\begin{figure}
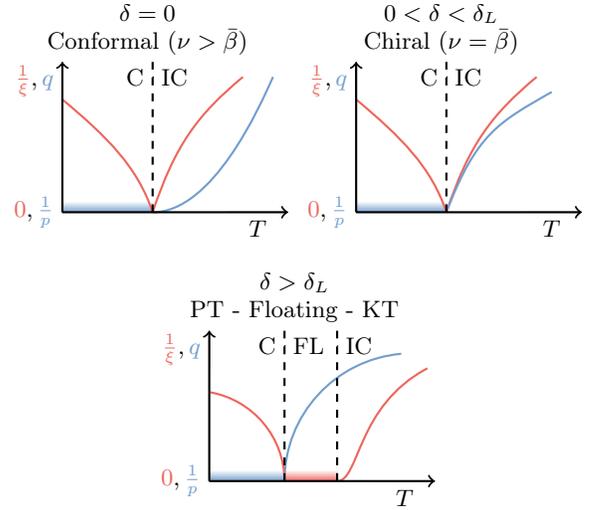

	\centering
	\[ \HuseFisher \]
	\caption{Sketch of the scenarios proposed by Huse and Fisher\cite{Huse.Fisher:1982} for commensurate-incommensurate (C-IC) phase transitions of the classical chiral clock model when the chiral perturbation $\delta$ is relevant. We define the $q$-vector in units of $2\pi$. In the case of the Rydberg model we will tune the Hamiltonian couplings instead of the temperature $T$ to probe for these transitions. The floating (FL) phase is also incommensurate but we distinguish it from the gapped incommensurate phase at higher temperatures.}
	\label{fig:Huse_Fisher}
\end{figure}

The most interesting cases are $p=3$ and $4$. Indeed, if the chiral perturbation $\delta$ is relevant, it was suggested\cite{Huse.Fisher:1982} that the IC-C transition may still be direct but in a new non-conformal (\textit{chiral}) universality class characterized by $\B = \nu$, at least up to a Lifshitz point $\delta_L$ beyond which the chiral perturbation becomes large enough for an intermediate floating phase to appear, and except possibly at isolated conformal points where the chiral perturbation vanishes. We present these three possible scenarios in Fig.~\ref{fig:Huse_Fisher}. As suggested by Huse and Fisher\cite{Huse.Fisher:1982}, the product $\xi|q-1/p|$ provides an accurate diagnosis since it is expected to diverge at the KT transition of a floating phase, to approach a strictly positive constant for a chiral transition, and to go to zero for a conformal transition.

Experimentally\cite{Bernien.Schwartz.ea:2017}, IC-C transitions were probed on a 1D system of 51 optically-trapped Rydberg atoms. The Kibble-Zurek (KZ) exponent $\mu$ is measured by dynamically tuning the laser parameters of the system and ramping through the IC-C transitions along specific lines. The experimental value of the KZ exponent, which controls the power-law increase of the domain size with the sweeping rate, is around $\mu\simeq 0.38$ for the $p=3$ case and $\mu\simeq0.25$ for $p=4$, while simulations reported a value around $\mu \simeq 0.45$ for $p=3$ and $\mu \simeq 0.2-0.3$ for $p=4$\cite{Keesling.Omran.ea:2019}. The agreement is good but not perfect, and these results call for further investigation of the nature of the phase transition. One major obstacle is the absence of exact results for the KZ exponent anywhere along the boundary that could serve as a benchmark. For infinite systems, the KZ exponent is related to the correlation length exponent $\nu$ and the dynamical exponent $z$ by the relation $\mu=\nu/(1+z\nu)$, but it has proved difficult to determine both $\nu$ and $z$ very accurately.
In this paper, we come up with numerically exact results for the KZ exponent across two points of the phase diagram of the experimental model\cite{Bernien.Schwartz.ea:2017}, one on the $p=3$ boundary, the other one on the $p=4$ boundary, by locating very accurately the lines in the IC phase where the $q$-vector is commensurate using the finite-size Density Matrix Renormalization Group\cite{White:1992} (DMRG) algorithm. Along these lines, since the system remains commensurate and chiral perturbations are absent, the transition, if it is unique and continuous, is expected to be conformal with dynamical exponent $z=1$.

For the $p=3$ case, the transition is expected to be in the universality class of the 3-state Potts (P) model, with $\nu=5/6$ and $\mu=5/11\simeq 0.4545$, while for the $p=4$ case it is expected to be Ashkin-Teller (AT)\cite{Ashkin.Teller:1943}, a family of universality classes parametrized by a coupling $\lambda$ and corresponding to two decoupled Ising models at $\lambda=0$ and to the symmetric 4-state Potts model at $\lambda=1$. Numerically, we have found an exponent $\nu\simeq 0.80$ corresponding to $\lambda\simeq 0.5$ and leading to $\mu\simeq 0.444$. These conformal points are located slightly below the tips of the corresponding lobes, and the value of the KZ exponent across these points can be used as benchmarks. Note that these values are significantly larger than those reported experimentally for 51 sites. Furthermore, the transition is found to be chiral in the vicinity of these points, with clear evidence that the product $\xi|q-1/p|$ neither vanishes nor diverges at the transition, and floating phases have been identified further away from the transition except below the period-3 phase.

This paper is structured as follows: In Sec.~\ref{sec:rydberg_model}, we introduce and review the experimentally relevant model of Rydberg atoms and its phase diagram. In Sec.~\ref{sec:dmrg_results} we present our main results and in Sec.~\ref{sec:discussion} we discuss our results in the context of past numerical and experimental results. Furthermore in Appendix~\ref{app:sec:method} we discuss the details of our particular DMRG implementation including the fitting of the measured correlation function, and in Appendix~\ref{app:sec:supplemental_data} we show additional results on complementary cuts to the ones shown in the main text as well as cuts going through the $p=2$ boundary, and also a finite size scaling analysis of the critical point drift.

\section{Rydberg model}
\label{sec:rydberg_model}

Experimentally\cite{Bernien.Schwartz.ea:2017}, each Rydberg atom of the chain can be excited to a Rydberg state by an applied laser with Rabi frequency $\Omega$ and detuning $\Delta$. Excited Rydberg atoms have long-range interactions between them, while they don't interact in the ground state. The hard-core boson Hamiltonian of this system is
\begin{equation}
	\Ha = \sum_{i} -\Delta \n_i + \Omega \s_i + \sum_{j > i} \frac{\n_i \n_j}{(i-j)^{6}},
	\label{eq:Rydberg}
\end{equation}
where $\n_i \equiv 2(\sigma^z+1)/2$  and $\s, \sigma^z$ are Pauli matrices. In the classical limit $\Omega = 0$, the repulsive interaction and the chemical potential compete. By adjusting their ratio, a devil's staircase\cite{Bak:1982, Nebendahl.Dur:2013} of classical ground states of many different ratios of occupation per unit cell size is generated, with the largest phases having one boson every $p$ sites. The $p$ phases are stable when $\Omega$ is turned on, up to values of $\Omega \sim \Delta$, beyond which the system becomes disordered.

\begin{figure}
	\centering
	\includegraphics{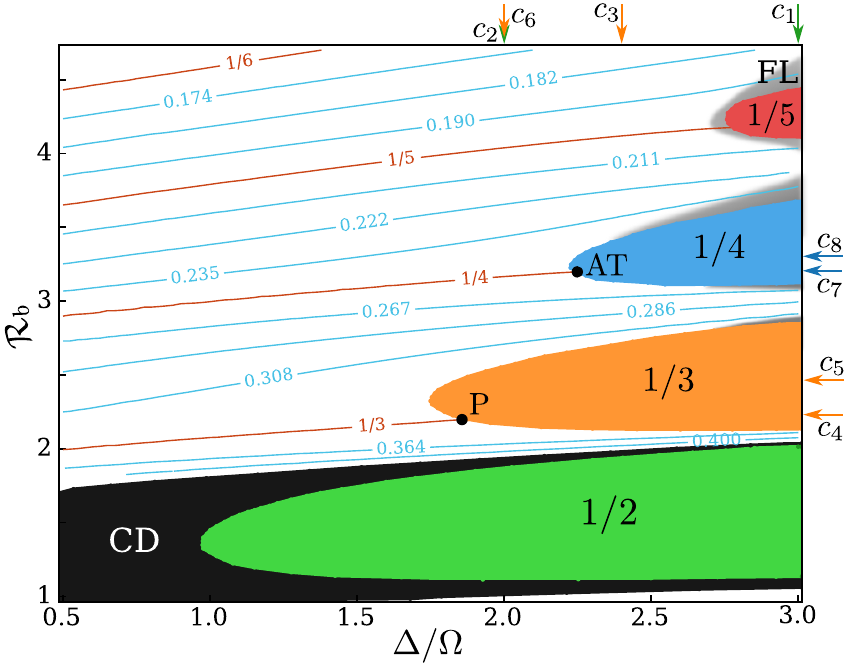}\put(-10, 85){}%

	\caption{Phase diagram of the Rydberg model obtained with DMRG simulations on 121 sites. Red, blue, orange and green regions are the ordered phases with period p=5, 4, 3, and 2 respectively, while the black region is a commensurate disordered (CD) phase with wave-vector $q=1/2$ in units of $2\pi$. The grey region is a sketch of the floating phase based on a previous iDMRG work\cite{Rader.Lauchli:2019}. Equal-$q$ lines are shown in the disordered and floating phases. The points P and AT are respectively our estimates of the Potts and Ashkin-Teller critical points. Apart from the cuts that go through these points, either horizontally/vertically, or along the associated commensurate lines (P/AT cuts), the other cuts discussed throughout the text are horizontal or vertical and are labeled $c_n$. They are represented by arrows colored according to the ordered phase they cross.}
	\label{fig:full_phase_diagram}
\end{figure}

The global phase diagram of the relevant region shown in Fig.~\ref{fig:full_phase_diagram} and plotted in the natural units $\x$ and $\y \equiv \Omega^{-1/6}$ has been obtained on chains of 121 sites. The $q$-vector has been deduced from a fit of the correlation function from the middle site with the Ornstein-Zernike form. For the sizes that we could reach with our finite-chain DMRG algorithm, it is not possible to map out the floating phase accurately, so we just show a sketch based on the results of a previous infinite DMRG study\cite{Rader.Lauchli:2019}, which for that matter is more accurate. Note that, according to that study, there is no floating phase around the tips of the period-4 and period-3 phase, and a floating phase was only observed beyond $\x \simeq 24$ below the period-3 phase.

\section{DMRG results}
\label{sec:dmrg_results}

\begin{figure*}
	\centering

	\includegraphics{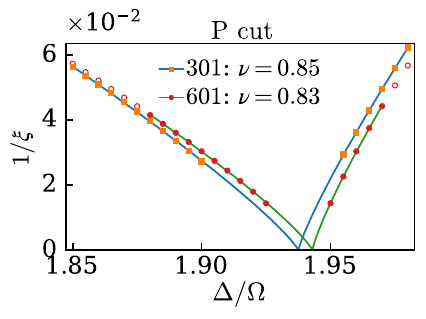}\put(-10, 85){(a)}%
	\includegraphics{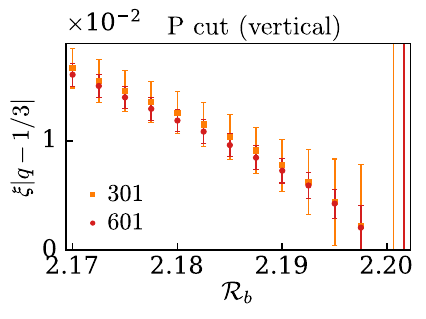}\put(-10, 85){(b)}%
	\includegraphics{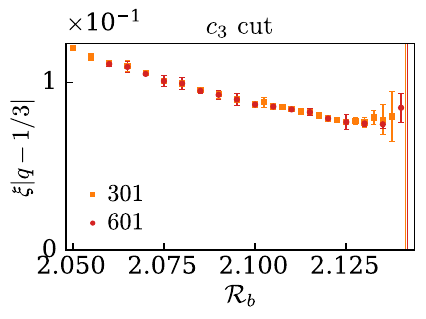}\put(-10, 85){(c)}%
	\includegraphics{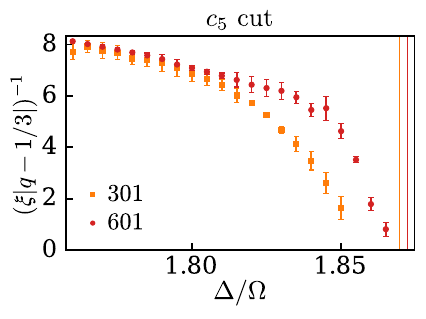}\put(-10, 85){(d)}%

	\includegraphics{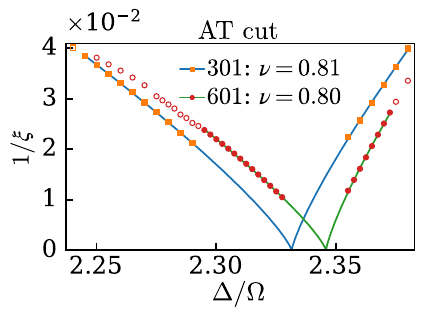}\put(-10, 85){(e)}%
	\includegraphics{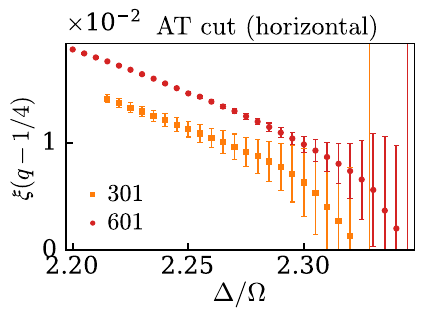}\put(-10, 85){(f)}%
	\includegraphics{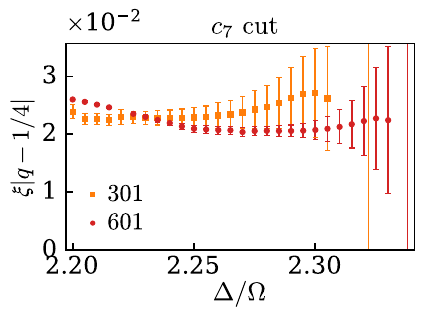}\put(-10, 85){(g)}%
	\includegraphics{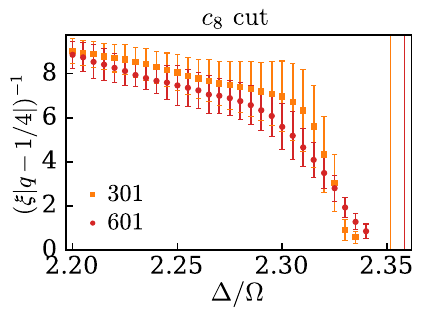}\put(-10, 85){(h)}

	\caption{Scaling across the boundaries of the period-3 (top) and period-4 (bottom) phases. Left panels: Inverse of the correlation length along the commensurate line with $q=1/3$ (top) and $1/4$ (bottom). Upon approaching the conformal 3-state Potts (P) and Ashkin-Teller (AT) points the product $\delta q\times \xi$ vanishes. In the vicinity of these points, along cuts $c_3$ and $c_7$, the product $\delta q\times \xi$ goes to a finite value, signalling a direct chiral transition in the Huse-Fisher universality class. Far away from the P and AT points, the product $\delta q\times \xi$ diverges (and its inverse goes to zero) at the Kosterlitz-Thouless transition.}
	\label{fig:main_panels}
\end{figure*}

To locate more accurately the conformal points, we have progressively refined the equal-$q$ lines in the vicinity of the period-3 and 4 lobes for 301 then 601 sites, reaching an accuracy in $q$ of the order of $10^{-4}$ (see Appendix~\ref{app:sub:equal_q_lines}). We then determined the correlation length $\xi$ along the $q=1/3$ and $1/4$ lines (P and AT cuts in Fig.~\ref{fig:main_panels}). The point where $\xi$ diverges, or equivalently where $1/\xi$ vanishes, is our estimate of the location of the conformal points, and the exponent with which it diverges is our numerical estimate of $\nu$.

To further characterize the conformal transitions, we considered vertical and horizontal cuts that go through the estimated P and AT points respectively. These are labeled "P/AT cut (vertical/horizontal)" in Fig.~\ref{fig:main_panels}. Along these cuts, $q$ varies, and accordingly one can estimate the exponent $\B$ and follow the behavior of the product $\xi|q-1/p|$. The two vertical lines on each of the plots of this product and the $\Delta q \equiv |q-1/p|$ plots are the $301$ and $601$ site estimates of the critical points obtained from the correlation length fit. To fit the $\Delta q$ power-laws, we fix the critical points to these estimates.

\begin{figure}
	\centering
	\includegraphics{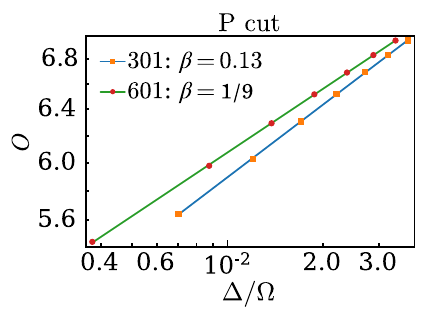}\put(-10, 85){(a)}%
	\includegraphics{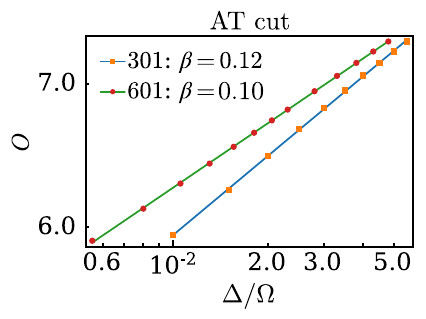}\put(-10, 85){(b)}
	\caption{Order parameter scaling along (a) the P and (b) the AT cuts (note the log-log scale). The data were fitted with the expected power-law behavior: $O \sim \|\x-(\x)_c\|^{\beta}$. For 301 sites both the exponent and the critical points are fitting parameters, while for 601 sites $\beta$ is fixed to the expected values to show the good agreement with our estimated $\nu$ along these cuts. The horizontal coordinates were shifted by the estimated critical point values.}
	\label{fig:O}
\end{figure}

\subsection{$q=1/3$ cut}
\label{par:_q_1_3_cut}

The commensurate $q=1/3$ line has been found to behave linearly close to the 1/3 phase, following approximately $\y \simeq 0.1284\x+1.9527$. Along this cut, we expect the transition to be 3-state Potts, with exponents $\nu = 5/6 \simeq 0.833$, $\B = 5/3 \simeq 1.66$ and $\beta = 1/9$\cite{Baxter:1980, Alexander:1975}. This point is called P in the phase diagram of Fig.~\ref{fig:full_phase_diagram}. It is located at $(\x, \y) \simeq (1.942, 2.202)$. Our results for $601$ sites agree within $1\%$ with the theory predictions.  The discrepancy in the $301$ site exponents could be due to finite-size effects that displace the $q=1/3$ line, so that along this cut we slightly missed the $301$ site equal-$q$ line. As expected, the concavities of $q$ and $\xi$ are opposite, and the product $\D$ converges to zero. Overall, our results provide strong evidence in favor of a 3-state Potts transition at point P.

\subsection{$q=1/4$ cut}
\label{par:_q_1_4_cut}

Turning now to the IC-C transition of the 1/4 phase, the $q=1/4$ equal-q line follows approximately $\y=0.1441\x+2.8747$ (AT cut) when we are very close to the $1/4$ phase. Along this commensurate line and for 601 sites, we find a phase transition at a point denoted AT in Fig.~\ref{fig:full_phase_diagram} at $(\x, \y) \simeq (2.346, 3.213)$ with exponent $\nu \simeq 0.80$, which is consistently replicated with a horizontal cut that crosses this critical point. The $\B$ exponent, unknown analytically for the AT universality class, is in any case larger than 1\cite{Huse.Fisher:1982, Huse.Fisher:1984}, and the $\D$ product decays to zero at the transition.

\subsection{Order parameter}
\label{par:order_parameter}

To further confirm the conformal nature of the transitions along the commensurate lines, we looked at the scaling of the order parameter $O$ defined as the maximal difference in the occupation $\braket{\n_l}$. To avoid the Friedel oscillations at the edges, we only consider the middle $10$ sites, leading to the following definition of $O$:
\begin{equation}
	O \equiv \max_{l\in J} \braket{\n_l} - \min_{l\in J} \braket{\n_l},
\end{equation}
where $J \equiv \{ \frac{L-1}{2} - 4, \dots, \frac{L-1}{2} + 5\}$ for odd $L$. The results are shown in Fig.~\ref{fig:O}. Along the P cut, and for 601 sites, the scaling is in excellent agreement with the exact result $\beta=1/9$. Along the AT cut, we expect the exponent $\beta$ to be related to $\nu$ by $\beta = \nu/8$, a prediction derived from the lowest CFT scaling dimension of the AT model\cite{Bridgeman.OBrien:2015}. For 601 sites, the scaling is in good agreement with $\beta=0.1$, the expected value for $\nu\simeq 0.8$.

\subsection{$p=3$ non-conformal cuts}
\label{par:p_3_non_conformal_cuts}

Let us now discuss the results we have obtained away from these points, starting with the period-3 phase. Both below the $1/3$ line ($c_3$ cut, vertical, $\x=2.4$) and above it ($c_4$ cut, horizontal, $\y=2.225$, Fig.~\ref{app:fig:third_2}) we find clear evidence of a chiral transition: $\D$ is nearly flat upon approaching the transition. Note that cut $c_3$ is remarkably far from the Potts point on the scale of the phase diagram of Fig.~\ref{fig:full_phase_diagram}, leaving a significant parameter range to probe the chiral universality class experimentally.

Along the cut $c_5$ at $\y=2.45$, further above, the IC-C transition is more consistent with Pokrovsky-Talapov, with $\nu=0.6$ and $\B\simeq 0.52$. It actually makes sense that $\B$ is more accurate since, with our algorithm, $q$ converges faster than $\xi$. On the disordered side, the correlation length grows rapidly until it eventually levels off before the PT transition. This is consistent with a KT transition into a floating phase, $\xi$ being limited by the finite size. The $\D$ product shows a clear divergence before $q$ becomes commensurate. This result might indicate that the floating phase reaches closer to the tip of the lobe than what is shown in the phase diagrams (see however Appendix~\ref{app:sub:finite_size_scaling} for a discussion of finite-size effects). The same conclusions apply to cut $c_6$ further above (Appendix~\ref{app:sub:period_3_and_4}).

\subsection{$p=4$ non-conformal cuts}
\label{par:_p_4_non_conformal_cuts}

The situation is very similar around the period-4 phase. The $c_7$ cut ($\y = 3.22$) is in agreement with a direct chiral transition of exponent $\nu$ ($\simeq \B$) slightly higher than at the AT point, suggesting that this exponent increases as we initially move away from the AT point. Cut $c_8$ ($\y = 3.32$) shows clear indication of an intermediate floating phase. Both exponents are in good agreement with the PT universality class, especially for 601 sites with $\nu'\simeq 0.52$ and $\B\simeq 0.47$, and the $\D$ product diverges at the KT transition, as expected.

\section{Discussion}
\label{sec:discussion}

\subsection{Comparison with blockade models}
\label{par:p_4_non_conformal_cuts}

It is instructive to compare these results with those obtained recently on \textit{blockade} models\cite{Fendley.Sengupta.ea:2004,Chepiga.Mila:2021*1}, in which configurations with bosons at a distance less or equal to $r=1, 2,\dots$ are forbidden while only the interaction at distance $r+1$ is kept, and which
are expected to be good effective models between the phases $p = r+1$ and $p=r+2$.

For the period-3 phase of the $r=1$ blockade model\cite{Fendley.Sengupta.ea:2004, Samajdar.Choi.ea:2018, Chepiga.Mila:2019}, there is a single point where the transition is conformal whose location is known exactly because it belongs to an integrable line\cite{Fendley.Sengupta.ea:2004}. Lines of chiral transitions seem to surround the Potts point\cite{Samajdar.Choi.ea:2018, Chepiga.Mila:2019}, while further away intermediate floating phases appear\cite{Fendley.Sengupta.ea:2004, Chepiga.Mila:2019}. Our results agree with all these properties. The only difference it that the chiral transition of our model is more extended below the lobe than for the blockade model. Note that, more generally, our results agree with those obtained on classical 2D and quantum 1D versions of the period-3 case\cite{Ostlund:1981*1, Huse:1981, Huse.Fisher:1982, Selke.Yeomans:1982, Huse.Szpilka.ea:1983, Schulz:1983, Howes.Kadanoff.ea:1983, Howes:1983, Huse.Fisher:1984, Selke:1984, Albertini.McCoy.ea:1989}, for which the existence of a transition line in the chiral universality class is supported both by experiments\cite{Abernathy.Song.ea:1994, Schreiner.Jacobi.ea:1994} and by recent numerical work\cite{Fendley.Sengupta.ea:2004, Samajdar.Choi.ea:2018, Whitsitt.Samajdar.ea:2018, Chepiga.Mila:2019, Nyckees.Colbois.ea:2021}.

In the context of Rydberg atoms, the $r=2$ blockade model has only been introduced and studied very recently\cite{Chepiga.Mila:2021*1}. The transition out of the period-4 phase along the commensurate line was found to be Ashkin-Teller with $\nu \simeq 0.78$ and $\lambda=0.57$. Our estimates $\nu\simeq 0.80$ and $\lambda=0.5$ are not far, confirming the qualitative relevance of blockade models. The parameter range of chiral transitions is comparable in both cases. Note that the presence of a range of chiral transition before a floating phase appears is in agreement with very recent results obtained on a classical 2D chiral Ashkin-Teller model\cite{Nyckees.Mila:2022}, according to which a chiral transition is expected for $\lambda \gtrsim 0.42$ and up to $\lambda \simeq 0.978$\cite{Schulz:1983}.

\subsection{KZ exponent}
\label{par:kz_exponent}

Finally, let us come back to the KZ exponent, and to the identification of the exponents of the chiral universality class, the main open issue in the field. On the theory side, the bottleneck is the determination of the dynamical exponent $z$. It is fixed to $z=1$ at the conformal points P and AT, but an accurate estimate of its value away from these points is still beyond state-of-the-art simulations. What one can get quite accurately however is the exponent $\nu$, and the fact that its value is consistent with that of $\B$ along the chiral transition is an indication that cross-over effects are negligible for the model of Rydberg chains, contrary to the classical 2D chiral Potts model, where crossover effects lead to an overestimate of $\nu$ and a violation of the $\nu=\B$ criterion close to the Potts point\cite{Nyckees.Colbois.ea:2021}.

On the experimental side, by contrast, one can accurately measure the KZ exponent. If measured on very large systems, this exponent should provide the missing piece of information on $\nu$ and $z$ since $\mu=\nu/(1+z\nu)$. How large should the systems be? The discrepancy between our numerically exact results at points P and AT and the experimental results on 51 sites clearly demonstrates that one needs larger systems. At point P, which corresponds to $\y\simeq 2.202$, the theoretical value is $\mu=5/11\simeq 0.4545$, while the experimental result is around $\mu\simeq0.38$. Similarly, at the AT point $\y\simeq 3.213$, our estimate is $\mu\simeq 0.444$, while the measured value is again much smaller, around $\mu\simeq 0.25$.

These remarks define a clear roadmap towards a conclusive identification of the chiral universality class with chains of Rydberg atoms. KZ experiments should be carried on across the conformal P and AT points identified in the present work on systems of increasing size until a quantitative agreement is reached with the numerically exact estimates of $\mu$ reported here. Then, a comparison between experimental values away from the conformal points and theoretical estimates of $\nu$ should allow one to reach precise conclusions regarding the critical exponents of the chiral transition. Work is in progress to refine our estimates of the exponent $\nu$ all along the boundary of the period-3 and period-4 phases where the transition is believed to be chiral. We hope that the present results will in parallel encourage experimentalists to perform KZ experiments on longer chains to help solve the long standing problem of the universality class of the chiral transition.

\begin{acknowledgments}
We thank Andreas Läuchli and Samuel Nyckees for useful discussions. This work has been supported by the Swiss National Science Foundation Grant No. 182179 and by the Delft Technology Fellowship (NC). The calculations have been performed using the facilities of the Scientific IT and Application Support Center of EPFL.
\end{acknowledgments}

\appendix

\section{Numerical method}
\label{app:sec:method}

\subsection{Algorithm}
\label{sub:algorithm}

We simulate the Rydberg model with our own 2-site DMRG\cite{White:1992} code in the Matrix Product State\cite{Dukelsky.Martin-Delgado.ea:1998} (MPS) formalism. We represent the Hamiltonian as a Matrix Product Operator (MPO) where the long-range power-law interaction is approximated by a sum of 12 exponentials\cite{Pirvu.Murg.ea:2010, Schollwock:2011}, which leads to an MPO virtual bond dimension of 14. The parameters of the exponential approximation are determined by a minimization of the cost function
\begin{equation}
	F \equiv \sum_{r=1}^{L} \left(r^{-6} - \sum_{i=1}^{12} u_i\lambda_i^{r}\right)^2.
\end{equation}
For all sizes considered, the minimized cost function was smaller than $10^{-16}$ (Table~\ref{app:tab:coefs}). As a comparison, a truncation of the power-law preserving the first 12 terms results in an equivalent squared differences error of $\sim 7.5 \times 10^{-14}$.

\begin{table}[t]
	\label{app:tab:coefs}
	\begin{tabular}{rr}
		\hline
		\multicolumn{1}{c}{$u_n$} & \multicolumn{1}{c}{$\lambda_n$} \\
		\hline
		\num{-4.375780e-2}        & \num{0.325988}                  \\
		\num{-3.707815e-2}        & \num{0.326220}                  \\
		\num{-1.216618e-6}        & \num{0.787502}                  \\
		\num{2.120790e-18}        & \num{1.013796}                  \\
		\num{4.620461e-7}         & \num{0.830014}                  \\
		\num{4.587174e-6}         & \num{0.730021}                  \\
		\num{1.936228e-05}        & \num{0.634314}                  \\
		\num{1.244534e-3}         & \num{0.429284}                  \\
		\num{6.849186e-2}         & \num{-0.017733}                 \\
		\num{7.054547e-2}         & \num{0.103117}                  \\
		\num{8.547828e-2}         & \num{0.322411}                  \\
		\num{8.550526e-1}         & \num{0.009142}                  \\
		\hline
	\end{tabular}
	\caption{Power-law fitting coefficients used for 601 sites, rounded to 6 decimal places. The resulting cost function $F$ is $3.8\times10^{-19}$. The closeness of some $\lambda_n$ suggest that the fit could be improved even further. However, it is not clear how to properly approach the search for a global minimum to the problem given the large number of parameters.}
\end{table}

To avoid stability problems in DMRG, we chose system sizes of the form $L = 12l + 1$, which split the ground state degeneracy by guaranteeing a single ground state with occupied edges for $p=3,4$. For the full phase diagram we chose $L=121$ which stabilizes all relevant orders.

On each two-site DMRG update where we carry out a singular value decomposition, we discard singular values smaller than $10^{-9}$, as these carry a statistical weight substantially smaller than machine precision, however, the truncation of the bond dimension to a hard limit $D$ is effectively much more relevant. Overall, the truncated weight in the last DMRG update in the middle of the chain was always lower than $10^{-6}$.

As a convergence criteria, we required the relative energy variance
\begin{equation}
	\frac{\braket{\Psi | H^2 | \Psi}}{\braket{\Psi | H | \Psi}^2} - 1,
\end{equation}
where $\ket{\Psi}$ is the variational MPS state, to be smaller than $10^{-11}$ when estimating the boundaries of phases and $10^{-12}$ when determining critical exponents and the $q=1/3, 1/4$ lines. An MPS virtual bond dimension of 350 was typically enough to reach such precision for $601$ sites and close to the $1/4$ conformal point, while bond dimensions up to 500 were used to reach convergence close to or inside the floating phases.

\subsection{Correlations and q-vector}
\label{app:sub:correlations_and_q_vector}

\begin{figure}
	\centering
	\includegraphics{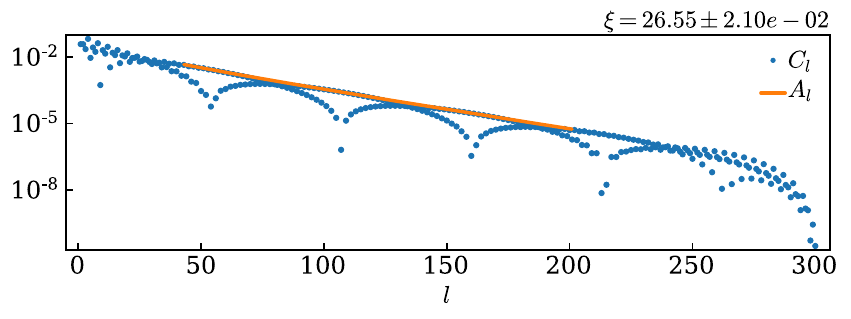}%
	\put(-12, 4){(a)}%

	\includegraphics{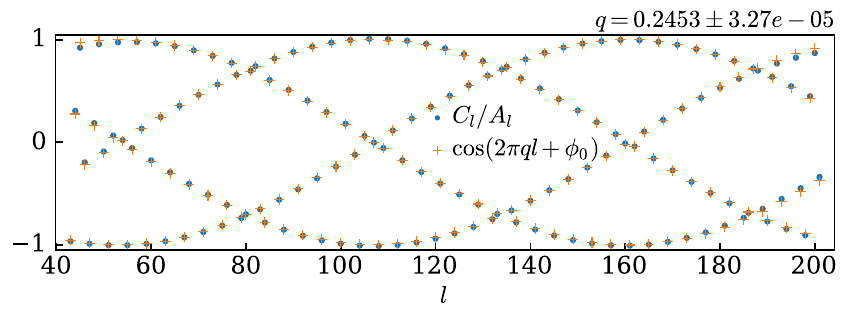}%
	\put(-12, 4){(b)}%
	\caption{Demonstration of the two-step fitting scheme for a 601 site simulation along the $c_8$ cut, at $(\x, \y) = (2.23, 3.32)$.}
	\label{app:fig:fits}
\end{figure}

We obtain $q$ by fitting the correlation function between the middle site $j$ and site $j+r$,
\begin{equation}
	C_{r} \equiv \braket{\n_{j}\n_{j+r}} - \braket{\n_{j}}\braket{\n_{j+r}},
\end{equation}
with the expected Ornstein-Zernike (OZ) form\cite{Ornstein.Zernike:1914}:
\begin{equation}
	\label{app:eq:OZ}
	C_r \sim A_r \cos{(2\pi q r + \phi_0)},
\end{equation}
where
\begin{equation}
	A_r \equiv A_0\frac{e^{-r/\xi}}{\sqrt{r}}.
\end{equation}
We discard points from the head and the tail-end of the correlation function until an OZ regime is thought to be reached, then we fit the remaining points. We implemented a two-step fitting scheme (Fig.~\ref{app:fig:fits}) that has been described before\cite{Chepiga.Mila:2021*1}, where first we obtain $\xi$ and $A_0$ by performing a linear fit on $C(r)\sqrt{r}$ in a semi-log scale. Then, $q$ is obtained by a least-squares cosine fit on $C(r)/A(r)$, where one minimizes the cost function $F(q)$ defined as the sum of squared differences. The confidence intervals (error bars) shown in the plots of $q$ are an estimate of the fitting error. They are calculated by assuming that the error $\delta q$ is proportional to the cost function. It then follows that $\delta q = F (dF/dq)^{-1}$ in lowest order, which can be explicitly calculated. The main contributions to this error are not precision errors in the fitting algorithm, but are instead errors in the determination of $\xi$ and $A_0$, or deviations from an OZ regime. The error bars of $\xi$ were deemed too small to be represented. The error of $\D$ is derived from these errors by the differential chain rule.

In general, we find the limit of reliable correlation lengths to be $\xi \sim L/6$, beyond which $\xi$ is noticeably limited by the finite size. Still, we find that the $q$-vector suffers less from finite size effects than $\xi$, and that a cosine fit beyond this $\xi$ limit can still give an accurate estimate of $q$.

\section{Supplemental Data}
\label{app:sec:supplemental_data}

\subsection{Period-2}
\label{app:sub:period_2}

\begin{figure}
	\includegraphics{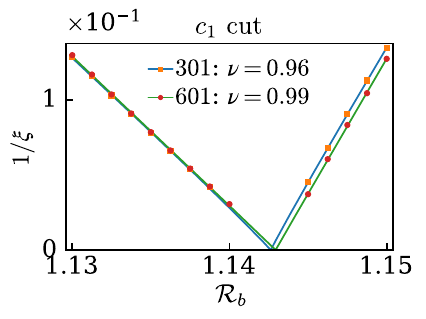}\put(-10, 85){(a)}%
	\includegraphics{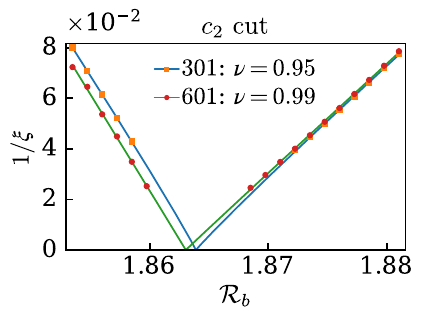}\put(-10, 85){(b)}%
	\caption{Correlation length along the $\x=3$ cut (left) that crosses the commensurate transition line below the period-2 lobe, and the $\x=2$ cut (right) that crosses it above. All points shown in the disordered sides are inside the commensurate phase. The small finite-size effects observed let us conclude from the exponent obtained that the transition is in the Ising universality class.}
	\label{app:fig:Ising}
\end{figure}

In contrast to the $p\geq3$ cases, the 1/2 lobe is surrounded by a commensurate disordered (CD) phase. The phase transition is continuous in the Ising universality class. We confirmed the latter by taking several cuts along the phase boundary and verifying that upon approaching the transition the correlation length diverges with the critical exponent $\nu \approx 1$. We confirm this at least up to the deepest cuts we considered at $\x=3$, as seen in Fig.~\ref{app:fig:Ising}. Above the lobe (on the side closer to the $1/3$ phase), we looked at cuts up to $\x=2$, with similar results. We did not look at cuts beyond $\x=2$ above since it was expected already from the effective $p=3$ blockade model that the transition would be Ising on this side\cite{Fendley.Sengupta.ea:2004, Chepiga.Mila:2019}. However, as we move away from the 1/3 phase, the Ising critical line of the blockade model eventually ends at a tricritical Ising point, below which the transition is first order. We did not find any evidence of a first order transition in the Rydberg model. It is not very surprising though because the tricritical point of the blockade model is located at negative (attractive) next-to-blockade interactions, which naturally does not occur in the Rydberg model. A more appropriate effective model of the lower part of the 1/2 lobe is the $p=2$ "blockade" hard-core boson model, which is the same as the Rydberg model in Eq.~(\ref{eq:Rydberg}) but where the interaction is truncated to the first term, a nearest neighbor interaction. A change of variables to a spin system reduces this model to an Ising model with transverse and longitudinal fields where the transition is always Ising\cite{Ovchinnikov:2003}.

\subsection{Equal-q lines}
\label{app:sub:equal_q_lines}

\begin{figure*}
	\centering
	\includegraphics{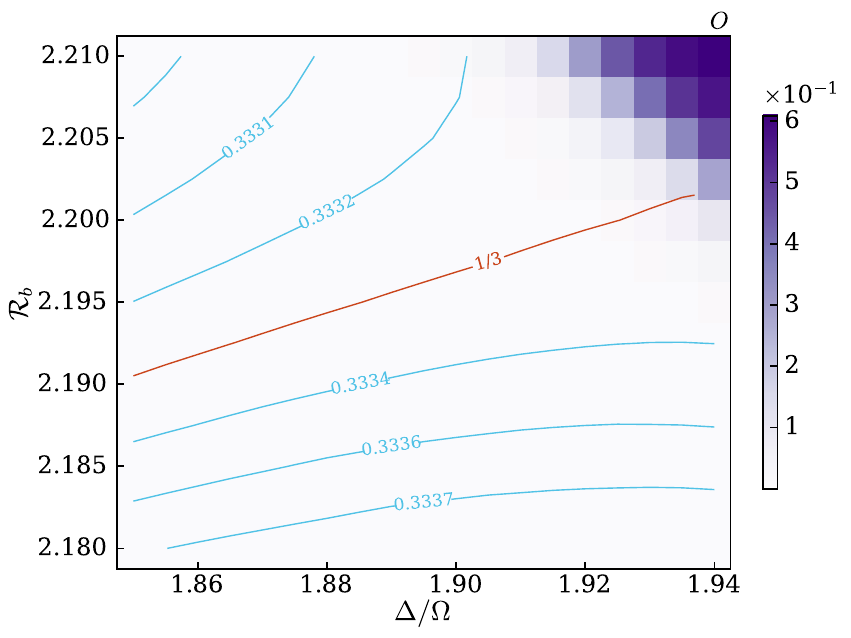}\put(-20, 175){(a)}%
	\includegraphics{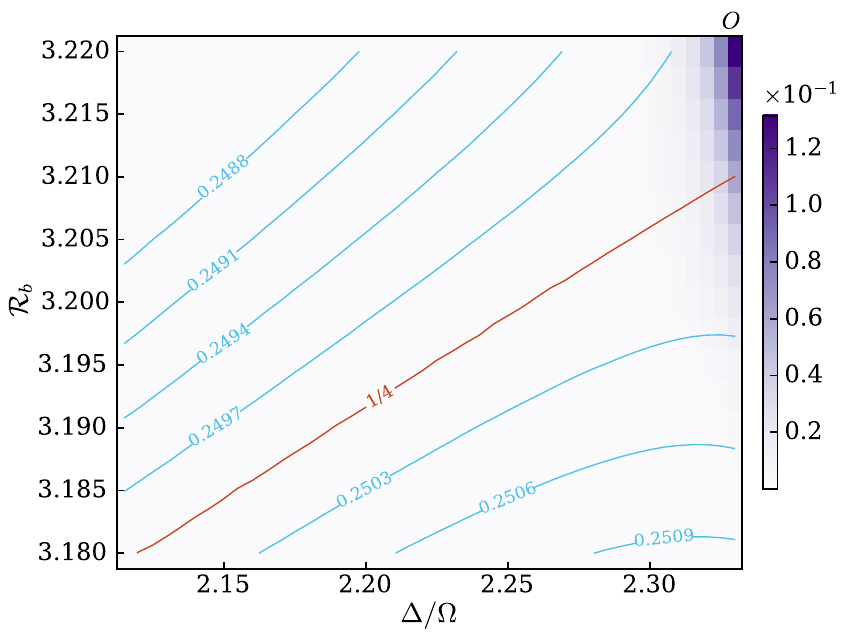}\put(-20, 175){(b)}%
	\caption{Regions where the $q=1/3,1/4$ lines meet their respective ordered phases. The colored grid data shows the order parameter $O$ (see main text). A linear fit of the $1/3$ line gives $\y=0.1284\x+1.9527$ (P cut), while for $1/4$ we have $\y=0.1441\x+2.8747$ (AT cut). All other equal-$q$ lines are repelled when approaching the ordered phases. }
	\label{app:fig:region}
\end{figure*}

The equal-q lines we show in the phase diagrams are obtained by interpolation of the $q$-vector on a finite grid. We use this same method to accurately determine where the $q=1/3$ and $1/4$ lines meet their respective ordered phases, using data from simulations on $601$ sites very close to the phase boundary, as shown in Fig.~\ref{app:fig:region}. The grid data in these figures show the order parameter $O$. We can see in these figures the start of the ordered phases in the top right. Simulations along these linear fits then lead to estimates of the conformal critical points.

\subsection{Period-3 and 4}
\label{app:sub:period_3_and_4}

\begin{figure*}
	\centering
	\includegraphics{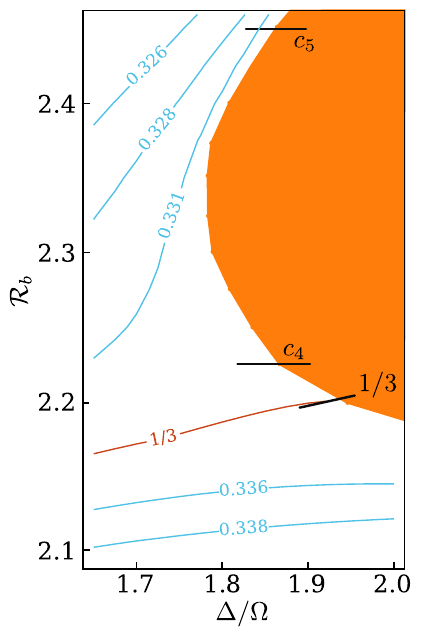}\put(-4, 178){(a)}%
	\begin{minipage}[b]{1.5\columnwidth}
		\includegraphics{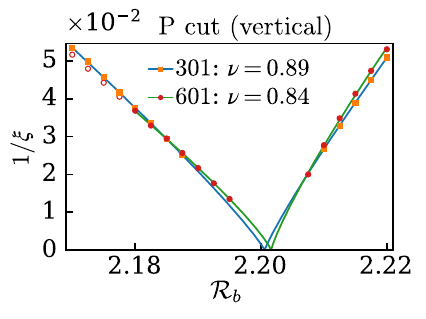}\put(-10, 85){(b)}%
		\includegraphics{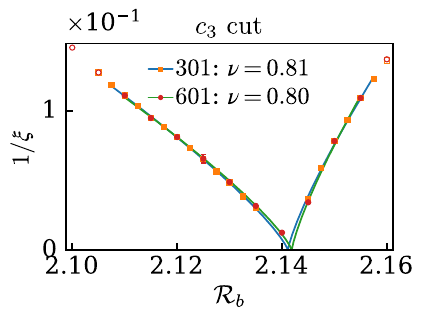}\put(-10, 85){(c)}%
		\includegraphics{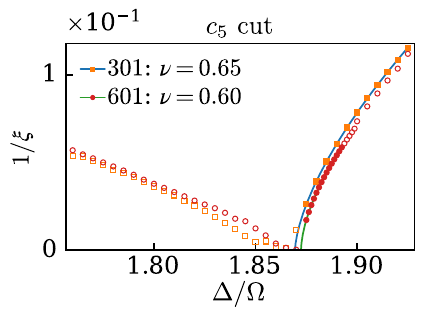}\put(-10, 85){(d)}%

		\includegraphics{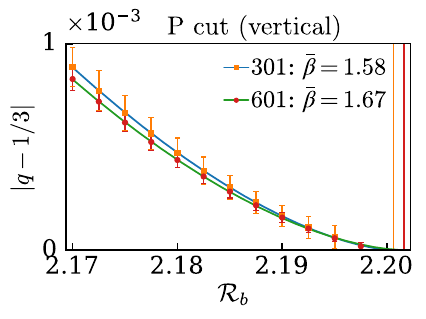}\put(-10, 85){(e)}%
		\includegraphics{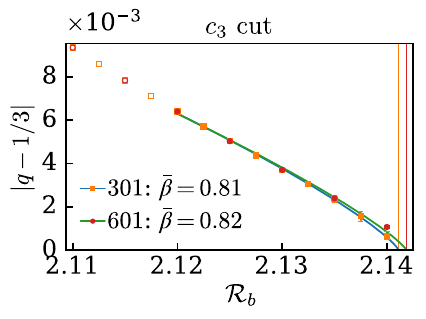}\put(-10, 85){(f)}%
		\includegraphics{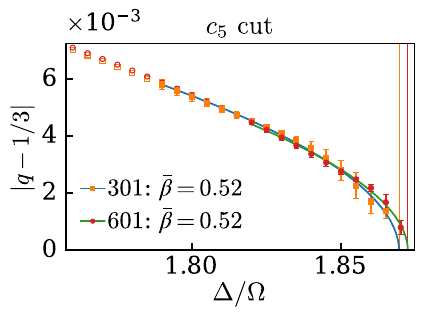}\put(-10, 85){(g)}%
	\end{minipage}
	\caption{Inverse of the correlation length (top) and distance of the incommensurate wave-vector $q$ to its commensurate value $1/3$ (bottom) for three cuts across the boundary of the period-3 lobe.}
	\label{app:fig:third}
\end{figure*}

\begin{figure*}
	\centering
	\includegraphics{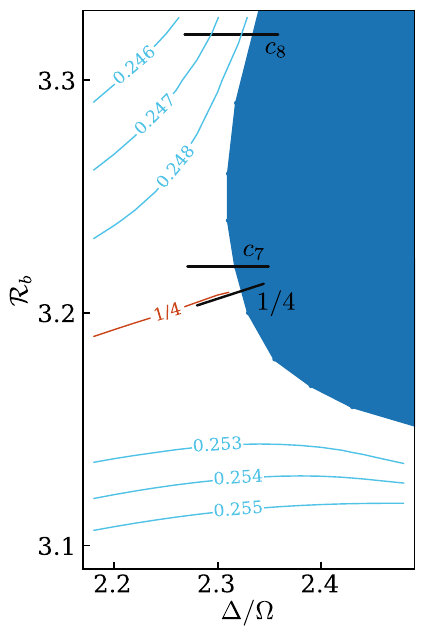}\put(-2, 178){(a)}%
	\begin{minipage}[b]{1.5\columnwidth}
		\includegraphics{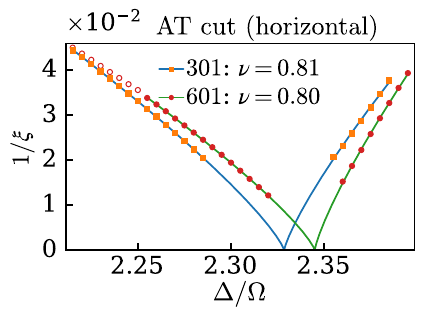}\put(-10, 85){(b)}%
		\includegraphics{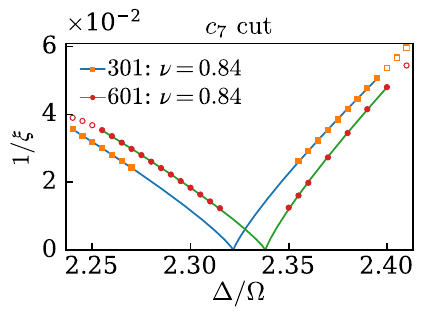}\put(-10, 85){(c)}%
		\includegraphics{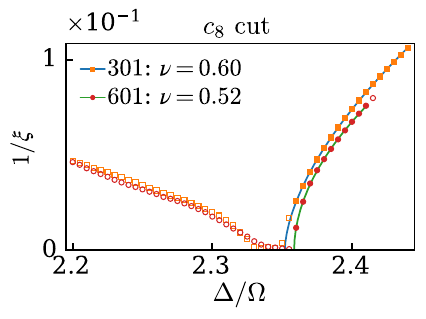}\put(-10, 85){(d)}%

		\includegraphics{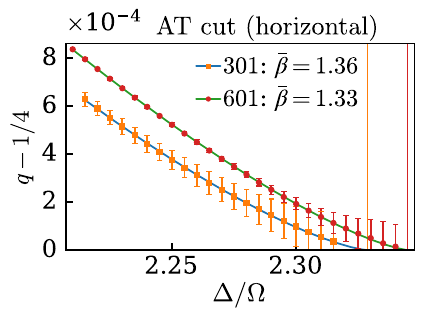}\put(-10, 85){(e)}%
		\includegraphics{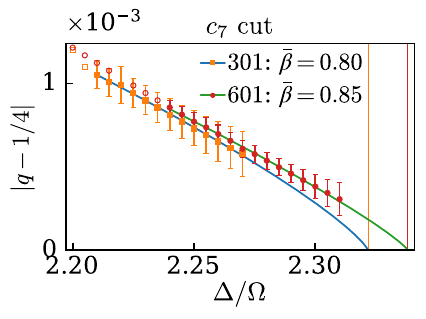}\put(-10, 85){(f)}%
		\includegraphics{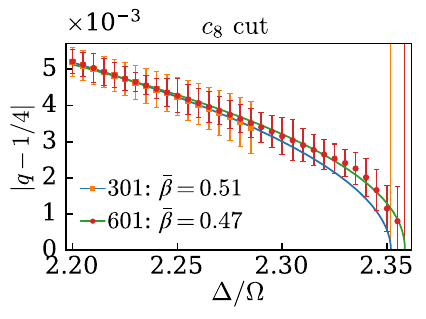}\put(-10, 85){(g)}%
	\end{minipage}
	\caption{Inverse of the correlation length (top) and distance of the incommensurate wave-vector $q$ to its commensurate value $\pi/2$ (bottom) for three cuts across the boundary of the period-4 lobe.}
	\label{app:fig:fourth}
\end{figure*}

In Figs.~\ref{app:fig:third} and~\ref{app:fig:fourth} we show the inverse $\xi$ and $q$-vector scaling along the incommensurate cuts crossing the $1/3$ and $1/4$ phase boundaries respectively, from which the $\D$ products shown in Fig.~\ref{fig:main_panels} of the main text have been obtained.

\begin{figure*}
	\centering
	\includegraphics{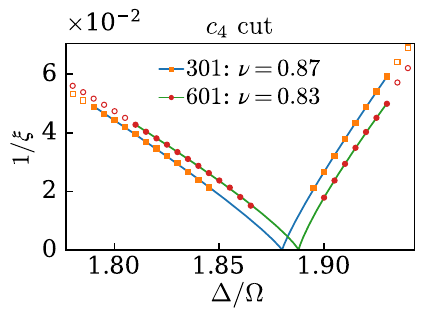}\put(-10, 85){(a)}%
	\includegraphics{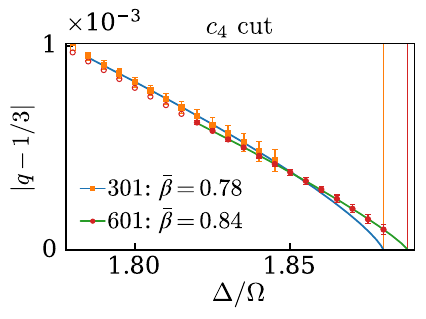}\put(-10, 85){(b)}%
	\includegraphics{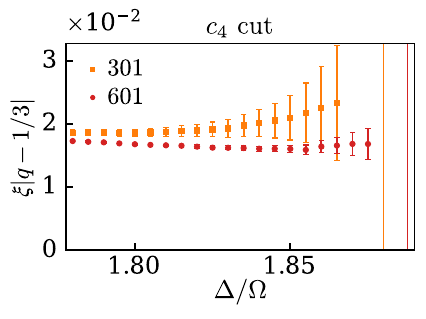}\put(-10, 85){(c)}%

	\includegraphics{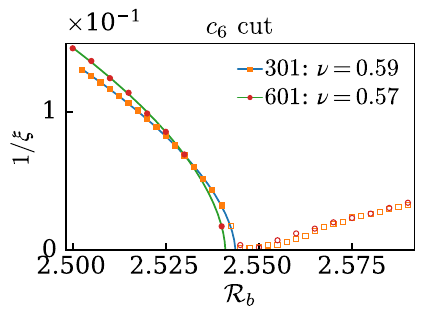}\put(-10, 85){(d)}%
	\includegraphics{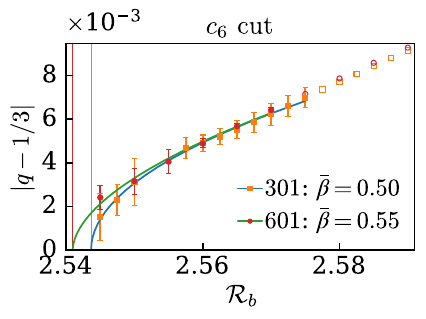}\put(-10, 85){(e)}%
	\includegraphics{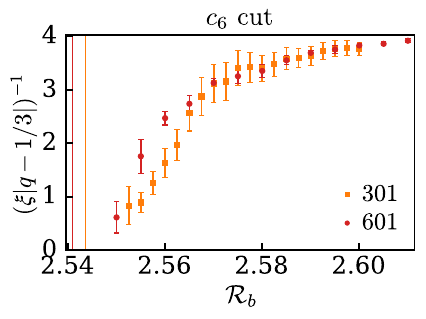}\put(-10, 85){(f)}%
	\caption{Inverse of the correlation length (left), distance of the incommensurate wave-vector $q$ to its commensurate value $1/3$ (middle) and the product $\D$ (right) for two additional cuts across the boundary of the period-3 lobe.}
	\label{app:fig:third_2}
\end{figure*}

We estimate the width of the floating phase by extrapolating to infinity the divergence of $\D$. For a cut $c_5$ we detect a floating phase of width in $\x$ of approximately $0.01$ for 301 sites and $0.004$ for 601 sites. As stated in the main text, this might be an indication that the floating phase reaches closer to the top of the lobe than what is shown in the phase diagrams. However, the shrinking width of the floating phase with system size might also suggest that there is a crossover to a chiral regime at larger system sizes. Similar reasoning can be applied to cut $c_6$ located further above in $\y$ and presented in Fig.\ref{app:fig:third_2} which turns out to be qualitatively equivalent to the cut $c_5$.

Fig.~\ref{app:fig:third_2} shows the results from two complementary cuts above the P point that were not included in the main text. The $c_4$ cut ($\y=2.225$) results are consistent with a chiral transition above but very close to the P point. Together with the $c_3$ cut, these two cuts suggest the P point is surrounded by chiral transition lines. The $c_6$ cut ($\x=2$) is qualitatively equivalent to the $c_5$ cut below it and brings further evidence in favor of an intermediate floating phase between the $1/3$ and $1/4$ phases.

\begin{figure*}
	\centering
	\includegraphics{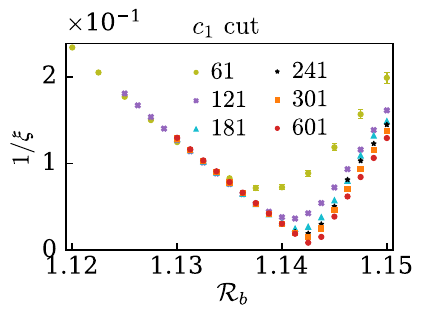}\put(-10, 85){(a)}%
	\includegraphics{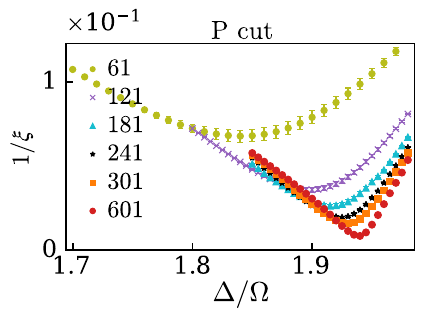}\put(-10, 85){(b)}%
	\includegraphics{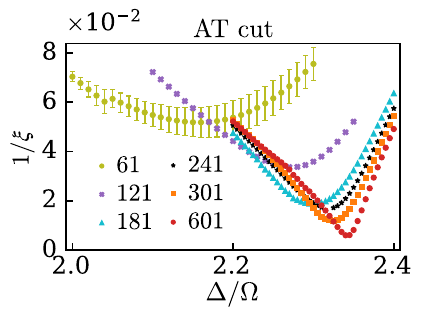}\put(-10, 85){(c)}%

	\includegraphics{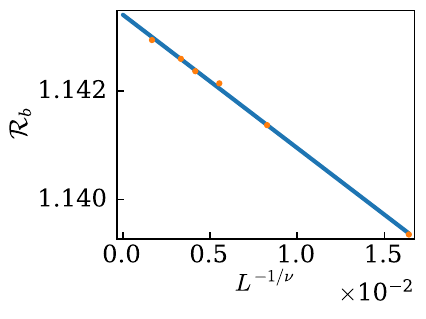}\put(-14, 93){(d)}%
	\includegraphics{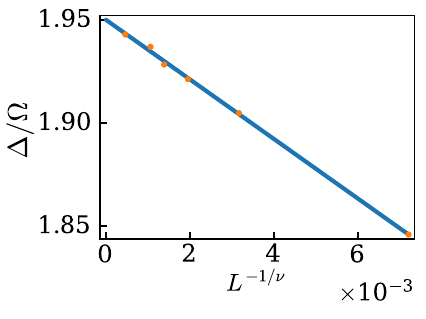}\put(-14, 93){(e)}%
	\includegraphics{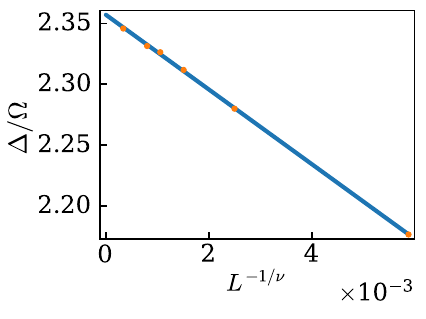}\put(-14, 93){(f)}%
	\caption{Finite size scaling analysis along the $c_1$ (Ising), P, and AT cuts. Top panels: inverse correlation length for several system sizes along the cuts. Bottom panels: finite-size scaling of the size-dependent critical points, where $\nu = 1, 3/5, 0.80$, for $c_1$, P, and AT respectively. The $L\rightarrow \infty$ limits lead to $\y = 1.1434$ for $c_1$, $\x = 1.951$ for P, $\x = 2.357$ for AT.}
	\label{app:fig:scaling}
\end{figure*}

\subsection{Finite-size scaling}
\label{app:sub:finite_size_scaling}

\begin{figure}
	\centering
	\includegraphics{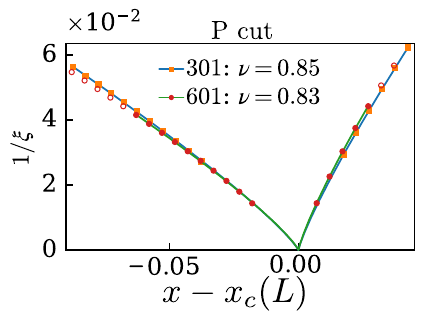}\put(-10, 85){(a)}%
	\includegraphics{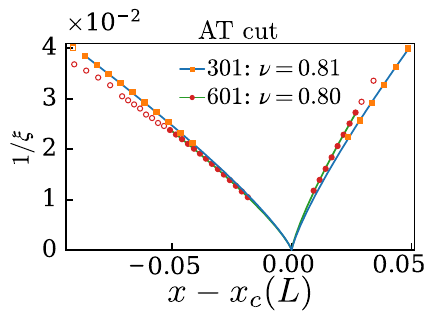}\put(-10, 85){(b)}%
	\caption{Inverse correlation length along the $p=3,4$ conformal cuts, as a function of the parametric distance to the estimated critical points ($x\equiv\Omega/\Delta$). }
	\label{app:fig:conformal_shift}
\end{figure}

It is already apparent from the results shown for 301 and 601 sites that a significant drift of the equal-$q$ lines and of the phase boundaries happens at small system sizes. Indeed, as shown in Fig.~\ref{app:fig:scaling}, the P and AT points show a significant drift between 601 sites down to experimentally relevant sizes like 61 sites. If we correct for the phase boundary drift, the finite-size difference in the correlation length between 301 and 601 sites is not as significant (Fig.~\ref{app:fig:conformal_shift}), although it is still more noticeable in the $p=4$ case. It's not unreasonable to expect a further drift of $\nu$ towards lower values for system sizes larger than we considered, possibly reaching closer to the blockade model prediction of $\nu \simeq 0.78$~\cite{Chepiga.Mila:2021*1}.

\bibliography{bib.bib}

\end{document}